\documentclass[aps,prl,superscriptaddress,twocolumn,showpacs]{revtex4}
\usepackage{amsmath}
\usepackage{graphicx}
\usepackage{amssymb}
\usepackage{bbm, color, soul}
\usepackage{graphicx}
\usepackage{adjustbox}

\usepackage{color}

\usepackage{tikz}

\usepackage{amsthm}
\usepackage{tikz-cd}
\usepackage[all]{xy}
\usepackage{amsfonts}
\usepackage{mathrsfs}
\usepackage{hyperref}
\usepackage{MnSymbol} 

\theoremstyle{plain}

\theoremstyle{definition}
\newtheorem{defn/}[equation]{Definition}

\theoremstyle{remark}
\newtheorem{rmk/}[equation]{Remark}
\newtheorem*{rmk*}{Remark}

\newtheorem*{claim*}{Claim}

\begin{document}

\title[]{Opposite amplitude–phase entropy responses at a non-Hermitian avoided crossing}

\author{Kyu-Won \surname{Park}}
\email{parkkw7777@gmail.com}
\affiliation{Department of Mathematics and Research Institute for Basic Sciences, Kyung Hee University, Seoul, 02447, Korea}

\author{Soojoon \surname{Lee}}
\email{level@khu.ac.kr}
\affiliation{Department of Mathematics and Research Institute for Basic Sciences, Kyung Hee University, Seoul, 02447, Korea}
\affiliation{School of Computational Sciences, Korea Institute for Advanced Study, Seoul 02455, Korea}

\author{Kabgyun \surname{Jeong}}
\email{kgjeong6@snu.ac.kr}
\affiliation{Research Institute of Mathematics, Seoul National University, Seoul 08826, Korea}
\affiliation{School of Computational Sciences, Korea Institute for Advanced Study, Seoul 02455, Korea}

\date{\today}

\begin{abstract}
Avoided crossings (A.C.) in open resonators arise from non-Hermitian mode interaction, where leakage produces complex
spectra and biorthogonal eigenmodes. Intensity-based entropies are robust markers of mode mixing but discard the phase
structure of the complex field. Here we introduce a field-level information-theoretic analysis based on the joint
statistics of local amplitude and phase under Born-weighted sampling on the cavity grid. For an open elliptical
microcavity in the strong-interaction A.C.\ regime, we find a distinctive sector-resolved response: amplitude statistics
tighten while phase statistics broaden maximally at the mixing point, and conditioning reveals strong amplitude–phase
dependence. By introducing a coarse position label and the associated co-information, we further show that the
enhancement of global amplitude–phase coupling is strongly shaped by spatial heterogeneity across the cavity.
\end{abstract}

\maketitle

\section{Introduction}
Open resonators and wave systems are intrinsically non-Hermitian: coupling to an external environment produces
irreversible leakage, complex resonance spectra, and biorthogonal eigenmodes
\cite{Rotter2009JPhysA,Moiseyev2011Book,Ashida2020AdvPhys}. Such effective non-Hermitian descriptions can be derived by
eliminating environmental degrees of freedom using projection-operator approaches \cite{Feshbach1958AnnPhys}. In this
setting, near-degeneracies are not only spectral features but also involve qualitative changes in eigenmode structure,
especially near exceptional points (EPs) where eigenvalues and eigenmodes coalesce
\cite{Kato1966Book,Heiss2012JPhysA,Berry2004CzJPhys}. Recent progress has further clarified how non-Hermiticity reshapes
spectral geometry and topology across platforms \cite{Bergholtz2021RMP,Ding2022NatRevPhys,OkumaSato2023ARCMP}.

Open dielectric microcavities offer a clean photonic platform for exploring such effects because radiative loss is
intrinsic and can be tuned continuously by shape deformation \cite{Vahala2004Book,Harayama2011LPOR,CaoWiersig2015RMP}.
In parallel, non-Hermitian photonics has rapidly expanded to include topological and device-level directions, supported
by recent reviews and tutorials \cite{Parto2020Nanophotonics,Nasari2023OME,Wang2023AOP}. Landmark experiments have
demonstrated exceptional rings, dynamical encircling, and microcavity sensing at EPs
\cite{Zhen2015Nature,Doppler2016Nature,Chen2017Nature}. Current trends emphasize practical constraints and noise-aware
interpretations of EP-based sensing and functionality \cite{Wiersig2020PRJ,Li2023NatNano,Meng2024APL}.

A widely used way to characterize mode interaction is through the intensity distribution and its real-space
delocalization. Across an avoided crossing (A.C.), intensity patterns typically hybridize and exchange modal character,
and intensity-based entropies provide robust signatures of this redistribution
\cite{Shannon1948BSTJ,Jaynes1957PhysRev,CoverThomas2006Book}. However, intensity is a gauge-invariant observable that
discards the phase structure of the complex eigenfield and is therefore insensitive to non-Hermitian features that
reside in the phase and in eigenmode non-orthogonality. This motivates a complementary, field-level diagnostic that
keeps the full complex eigenmode.

In this work we develop an information-theoretic framework that treats the local amplitude and phase of the complex
field as statistical variables induced by intensity-weighted sampling on the cavity grid. This construction allows us
to track how amplitude and phase statistics reshape across the A.C.\ and to quantify their dependence through marginal
and conditional measures, using standard information-theoretic tools and estimators
\cite{MacKay2003Book,CoverThomas2006Book,Kraskov2004PRE}. Applying the framework to an open elliptical microcavity, we
identify a distinctive signature of the A.C.\ that cannot be inferred from intensity alone: the amplitude statistics
tighten at the mixing point, while the phase statistics broaden maximally, and conditioning shows that amplitude and
phase remain strongly coupled in the interaction window. The resonant fields used for our analysis are computed for open
dielectric cavities using boundary-integral and boundary-element techniques \cite{Wiersig03_BEM}.

To further distinguish dependence that is genuinely local from dependence enhanced by spatial mixing across the cavity,
we introduce an additional random variable given by a coarse position label obtained by partitioning the cavity into
spatial bins. This enables a position-conditioned characterization and the use of co-information (interaction
information) as a compact measure of how spatial heterogeneity mediates the observed global amplitude--phase coupling
\cite{McGill1954Psychometrika}.

The paper is organized as follows. We first introduce the open elliptical microcavity and the spectral signatures of
the A.C.\ in the strong-interaction regime. We then define the Born-weighted amplitude--phase probability model and the
associated information measures. Next we present the amplitude and phase entropy signatures across the A.C., and
finally we introduce the coarse position label and co-information to disentangle within-region dependence from spatial
heterogeneity effects.


\section{Non-Hermitian mode interaction and avoided crossing in an open ellipse}

\subsection{Non-Hermitian effective description and strong-interaction regime}
Open wave systems are naturally described by non-Hermitian operators: radiative leakage and irreversible decay render
the resonance spectrum complex, and the associated eigenmodes biorthogonal~\cite{Rotter2009JPhysA,Moiseyev2011Book}. In
the vicinity of an exceptional point (EP), where both eigenvalues and eigenmodes coalesce, non-Hermitian effects become
particularly pronounced and control the qualitative transition between weak and strong mode interaction~\cite{Kato1966Book,Heiss2012JPhysA}.
An open dielectric microcavity provides a convenient photonic platform to access this physics, because the leakage is
intrinsic and can be tuned continuously by shape deformation~\cite{CaoWiersig2015RMP}.

Eliminating the bath degrees of freedom leads to an effective non-Hermitian Hamiltonian of the form
\begin{equation}
H \;=\; H_S \;+\; V_{SB}\,G_{B}^{(\mathrm{out})}\,V_{BS},
\label{eq:Heff_Green}
\end{equation}
where $H_S$ is the Hermitian operator of the corresponding closed system, $G_{B}^{(\mathrm{out})}$ is the outgoing Green
function in the bath, and $V_{SB}$ ($V_{BS}$) describes the system-to-bath (bath-to-system) coupling. To connect this
framework to the experimentally relevant spectral signatures, we use a minimal two-mode reduction parametrized by the
eccentricity $e$,
\begin{equation}
H(e)=
\begin{pmatrix}
\varepsilon_1(e) & v\\
v & \varepsilon_2(e)
\end{pmatrix},
\qquad
\varepsilon_j(e)=\nu_j(e)-i\,\gamma_j(e),
\label{eq:Heff_2mode_e}
\end{equation}
with real coherent coupling $v\in\mathbb{R}$ and loss rates $\gamma_j\ge0$. The eigenvalues are complex and can be
written as
\begin{equation}
E_{\pm}(e)=\frac{\varepsilon_1(e)+\varepsilon_2(e)}{2}\ \pm\ Z(e),
\label{eq:eigs_2mode_e}
\end{equation}
where $Z(e)=\sqrt{\left[\left(\varepsilon_1(e)-\varepsilon_2(e)\right)/2\right]^2+v^2}$.
In our microcavity setting, these complex eigenvalues correspond to the complex resonant wavenumbers
$k=\Re(k)-i\,\Im(k)$ of the Helmholtz problem: $\Re(k)$ determines the resonance frequency while $\Im(k)>0$ encodes the
radiative loss rate.

Near resonance, the interaction regime is governed by the competition between coherent coupling and differential loss.
One obtains a repulsion in $\Re(E_{\pm})$ with a crossing in $\Im(E_{\pm})$ for
$2v>|\Im(\varepsilon_1)-\Im(\varepsilon_2)|$ (strong interaction), whereas $\Im(E_{\pm})$ repel with $\Re(E_{\pm})$
crossing for $2v<|\Im(\varepsilon_1)-\Im(\varepsilon_2)|$ (weak interaction). The boundary $Z(e)=0$ corresponds to an EP
and marks the transition between these two regimes. In our data, the observed spectral pattern is the strong-interaction
case: a clear avoided crossing in $\Re(k)$ accompanied by a crossing in $\Im(k)$ within the eccentricity window analyzed
below. We note that the complementary weak-interaction case yields similar qualitative conclusions for the field-based
information diagnostics developed in the following sections.

\subsection{Open elliptical microcavity and spectral signatures}
Having established the minimal non-Hermitian framework, we now specify the concrete wave system used
throughout the paper and summarize the spectral signatures that identify the avoided crossing (A.C.)
pair.

We study two-dimensional dielectric microcavities governed by the scalar Helmholtz equation for TM
polarization,
\begin{align}
(\nabla^2 + n^{2}k^2)\,\psi(\mathbf r) = 0,
\label{eq:HelmholtzOpen}
\end{align}
where $\psi(\mathbf r)$ denotes the out-of-plane electric field ($E_z$) at position $\mathbf r=(x,y)$.
The refractive index is piecewise constant, $n(\mathbf r)=n_{\rm in}=3.3$ inside the cavity and
$n(\mathbf r)=n_{\rm out}=1$ outside. At the dielectric boundary we impose the standard TM interface
conditions, i.e., continuity of $\psi$ and of the normal derivative $\partial\psi/\partial n$, and in
the exterior we enforce the outgoing (Sommerfeld) radiation condition. As a result, the resonant
wavenumbers are complex, $k=\Re(k)-i\,\Im(k)$ with $\Im(k)>0$ encoding radiative loss. We compute
resonances and associated fields using a boundary element method (BEM), a boundary-integral scheme with
outgoing Green's functions that naturally yields a non-Hermitian effective operator for open
systems~\cite{Wiersig03_BEM}.

\begin{figure}
\centering
\includegraphics[width=9.5cm]{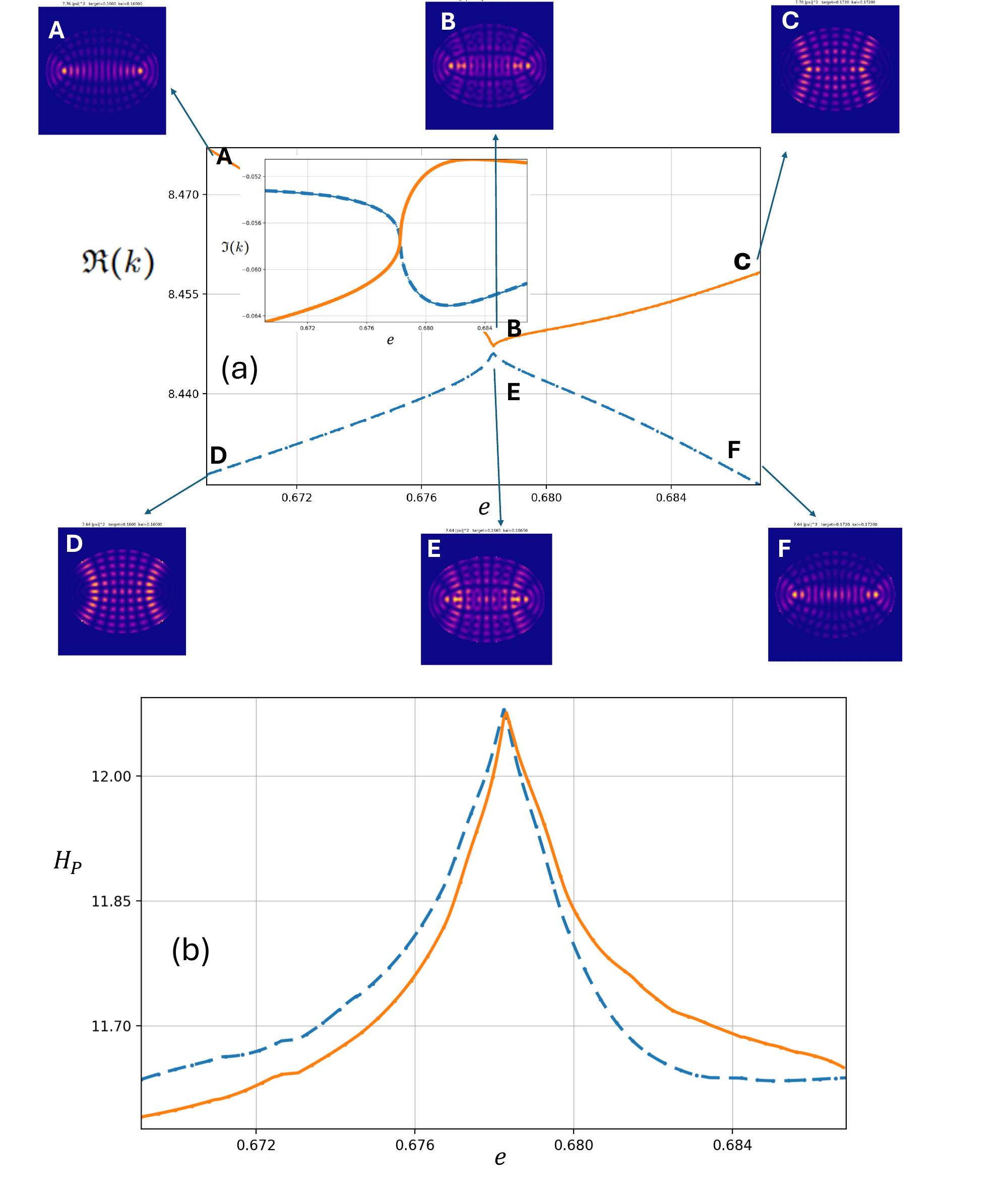}
\caption{\textbf{Avoided crossing and intensity-based spatial entropy.}
(a) $\Re(k)$ of two interacting resonances versus eccentricity $e$ (solid: mode~1; dashed: mode~2) shows a clear avoided
crossing; the inset plots $\Im(k)$ in the same window. Intensity snapshots A--F illustrate strong hybridization near the
A.C.\ (B,E) and exchange of modal character across it.
(b) Shannon entropy $H_P$ (bits) of the normalized spatial density
$p(\mathbf r)=|\psi(\mathbf r)|^2/\sum_{\mathbf r\in\Omega}|\psi(\mathbf r)|^2$ (cavity interior $\Omega$) peaks near the
A.C., quantifying enhanced delocalization of the \emph{intensity} distribution.}
\label{Figure-1}
\end{figure}

Throughout this work, the cavity boundary is taken to be an ellipse with semi-axes
$a=(1+\varepsilon)$ and $b=(1+\varepsilon)^{-1}$, so that the area $\pi a b$ is fixed to $\pi$ while the
shape is controlled by the deformation parameter $\varepsilon$ (with $a$ the major and $b$ the minor
semi-axis). For later convenience we also introduce the corresponding eccentricity
$e=\sqrt{1-b^2/a^2}$ $(0\le e<1)$, which provides a one-to-one measure of the same shape deformation. In
the present dataset we focus on the interval $e\simeq 0.669$--$0.687$, corresponding to the deformation
window $\varepsilon\in[0.16,0.173]$, which contains the avoided-crossing region analyzed in
Fig.~\ref{Figure-1}.

Figure~\ref{Figure-1}(a) shows two interacting eigenmodes forming an A.C.\ pair: as $e$ is varied, the two
branches in $\Re(k)$ approach and repel, while the inset confirms a concomitant variation in $\Im(k)$.
The six intensity snapshots A--F visualize the real-space manifestation of this coupling through
$|\psi(\mathbf r)|^2$. Far from the interaction window the two branches exhibit distinct intensity
morphologies, whereas near the A.C.\ (B and E) the patterns become strongly hybridized. Crossing the
A.C., the modal character is exchanged between the branches, consistent with the spectral repulsion in
$\Re(k)$ and the correlated response of $\Im(k)$.

To quantify the spatial redistribution associated with this spectral interaction, Fig.~\ref{Figure-1}(b)
plots the classical Shannon entropy $H_P$ of the normalized spatial probability density
$p(\mathbf r)=|\psi(\mathbf r)|^2\big/\sum_{\mathbf r\in\Omega}|\psi(\mathbf r)|^2$, where $\mathbf r=(x,y)$ denotes
the position on the cavity grid. By construction, $p(\mathbf r)\ge 0$ and it satisfies the normalization
$\sum_{\mathbf r\in\Omega}p(\mathbf r)=1$ over the cavity-interior grid points $\Omega$.
The entropy is evaluated in bits as
\begin{equation}
H_P=-\sum_{\mathbf r\in\Omega} p(\mathbf r)\log_2 p(\mathbf r),
\end{equation}
where $\Omega$ denotes the cavity-interior grid points.
Similar Shannon-entropy diagnostics of spatial mode redistribution have been used to characterize avoided crossings and
exceptional points in open microcavities~\cite{Park2018PRE,Park2020SciRepEP}, and have also been applied in other
platforms such as confined atomic systems and acoustic superlattices~\cite{SahaJose2020IJQC,SanchezDehesa2022FrontPhys, Nascimento2021EPJD}.

In this context, $H_P$ measures how broadly the intensity $|\psi(\mathbf r)|^2$ is spread over the cavity: larger $H_P$
indicates stronger spatial delocalization of the mode intensity. Notably, $H_P$ exhibits a pronounced peak in the
A.C.\ region, aligning with the strong hybridization observed in panels B and E. Away from the A.C., $H_P$ decreases as
the intensity recovers a more structured and localized pattern, indicating that the entropy enhancement is tightly
linked to mode mixing and the exchange of modal character across the avoided crossing.

\begin{figure*}[htbp]
\centering
\includegraphics[width=15.5cm]{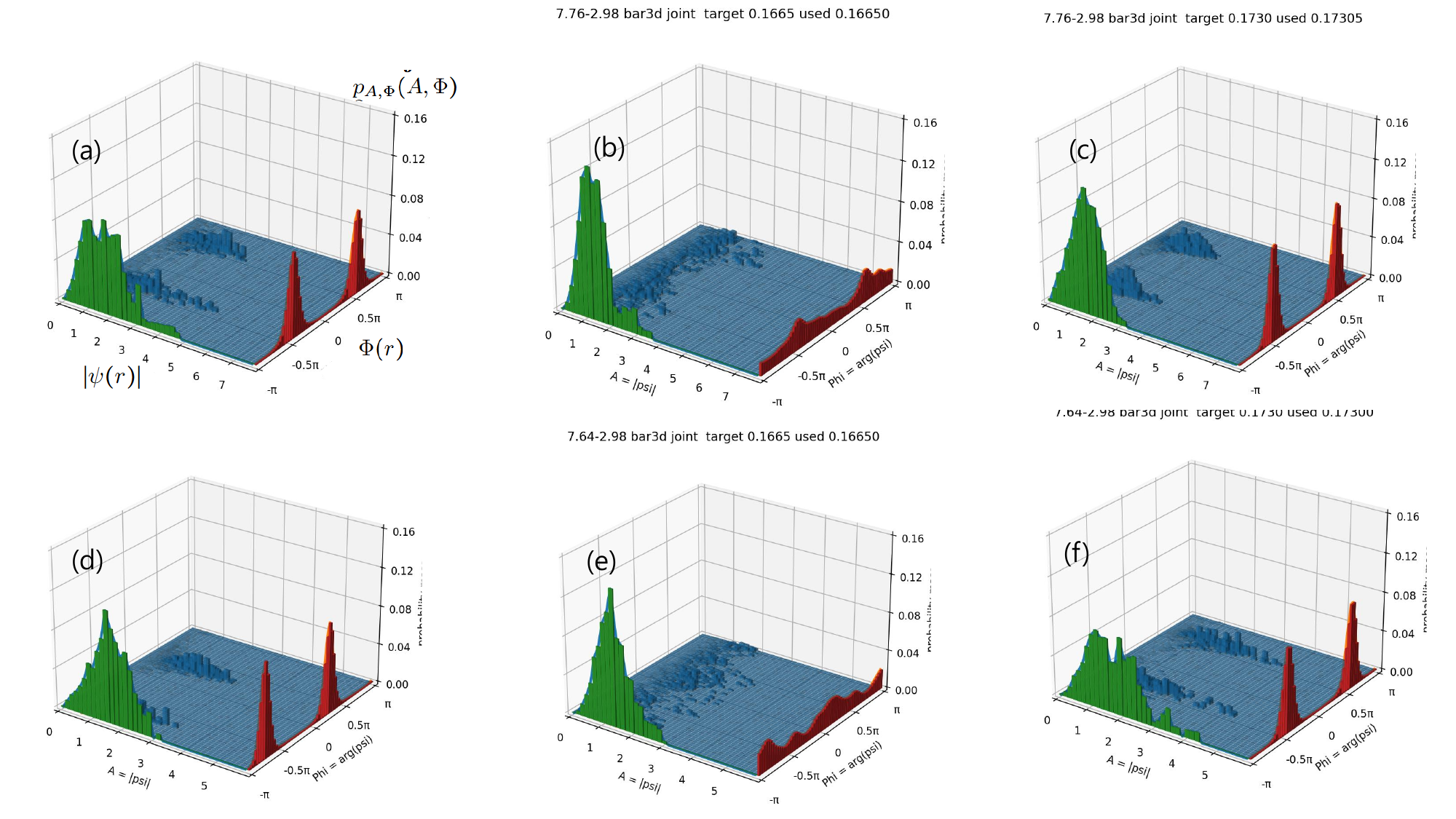}
\caption{\textbf{Born-weighted amplitude--phase joint distributions across the A.C.}
Joint histograms show $p_{A,\Phi}(A,\Phi)$ from cavity-interior points weighted by $w(\mathbf r)=|\psi(\mathbf r)|^2$,
with $A(\mathbf r)=|\psi(\mathbf r)|$ and $\Phi(\mathbf r)=\arg\psi(\mathbf r)\in[-\pi,\pi)$. Green/red walls are the
marginals $p_A(A)$ and $p_\Phi(\Phi)$. Panels (a--c) follow points A--C on the upper branch and (d--f) points D--F on
the lower branch in Fig.~\ref{Figure-1}(a), with B and E at the A.C. At the A.C.\ (b,e), $p_A$ narrows while $p_\Phi$
broadens, anticipating a local minimum of $H(A)$, an increase of $H(\Phi)$, and a peak of $I(A;\Phi)$.}
\label{Figure-2}
\end{figure*}
The intensity-based entropy in Fig.~\ref{Figure-1}(b) provides a coarse but robust marker of strong mode
interaction in real space. While $|\psi(\mathbf r)|^2$ is a gauge-invariant observable and thus experimentally
robust, it discards the phase structure and the non-orthogonality (biorthogonality) that are intrinsic to
non-Hermitian eigenmodes; these features become crucial precisely in the strong-interaction window near an EP.
This motivates us to probe the complex field $\psi(\mathbf r)$ itself through the joint statistics of its local
amplitude $A(\mathbf r)=|\psi(\mathbf r)|$ and phase $\Phi(\mathbf r)=\arg\psi(\mathbf r)$, and to quantify their
coupling via amplitude–phase information measures introduced next.

\section{Amplitude--phase probability model and information measures}
\subsection{Born-weighted sampling and random variables on the cavity grid}

While the spatial entropy $H_P$ captures the redistribution of the \emph{intensity} $|\psi(\mathbf r)|^2$, openness
renders the eigenmodes intrinsically non-Hermitian objects: they are generally biorthogonal and can exhibit reduced
phase rigidity, especially near mode interaction and EP-related regimes~\cite{Rotter2009JPhysA,Moiseyev2011Book,Heiss2012JPhysA}.
This means that diagnostics based solely on $|\psi(\mathbf r)|^2$ do not fully characterize the non-Hermitian eigenmode
structure. Motivated by this, we analyze the complex \emph{field} $\psi(\mathbf r)$ at the level of its local amplitude
$A(\mathbf r)=|\psi(\mathbf r)|$ and phase $\Phi(\mathbf r)=\arg\psi(\mathbf r)$, and quantify how their joint statistics
evolve across the avoided crossing. Such amplitude--phase information measures provide a complementary handle on
non-Hermitian mode mixing and offer a route toward diagnosing eigenmode properties beyond intensity-based observables.
Related information-theoretic diagnostics of complex eigenmodes across avoided crossings in open microcavities have also
been developed from complementary representations~\cite{Park2025PLA_ComplexEigenmodes}.

At this point it is important to clarify why we work with the amplitude--phase pair $(A,\Phi)$ rather than the Cartesian
decomposition $\psi(\mathbf r)=\Re\psi(\mathbf r)+i\,\Im\psi(\mathbf r)$. The latter is not invariant under a global
$U(1)$ gauge rotation $\psi(\mathbf r)\mapsto e^{i\theta}\psi(\mathbf r)$: a change of the arbitrary overall phase mixes
$\Re\psi$ and $\Im\psi$ through a rigid rotation in the complex plane, and thus reshuffles any statistics built directly
from the two components~\cite{Park2025PLA_ComplexEigenmodes}. In contrast, the amplitude $A(\mathbf r)$ is strictly
gauge-invariant, and the phase $\Phi(\mathbf r)$ transforms by a uniform shift $\Phi(\mathbf r)\mapsto \Phi(\mathbf r)+\theta$,
so that physically meaningful phase diagnostics can be formulated either in terms of relative phases or circular
statistics insensitive to the choice of global phase reference. This gauge-separation makes $(A,\Phi)$ a more robust
representation for characterizing non-Hermitian eigenmodes, whose overall phase is generically unconstrained and can vary
across parameter sweeps.

We now formalize the amplitude--phase statistics used in Figs.~\ref{Figure-2} and~\ref{Figure-3}. Unlike the spatial
entropy $H_P$, which depends only on the normalized density $p(\mathbf r)$, the present analysis treats the local field
observables as \emph{random variables} induced by Born-weighted sampling on the cavity grid~\cite{CoverThomas2006Book,MacKay2003Book}.
For a given eigenmode, the \emph{sample space} is the discrete set of cavity-interior grid points
$\Omega=\{\mathbf r_1,\dots,\mathbf r_N\}$ with $\mathbf r=(x,y)$, and a single \emph{outcome} is the selection of one
grid point $\mathbf r\in\Omega$. Here $\Omega$ includes only cavity-interior grid points; exterior points by construction
are excluded, while interior nodal points (zeros within the cavity) are retained. Throughout, grid points are sampled
according to the normalized Born weight $P(\mathbf r)\propto |\psi(\mathbf r)|^2$ on $\Omega$.

The amplitude and phase are then random variables on $(\Omega,P)$, namely maps from the sample space of grid points to
their respective \emph{value spaces} (codomains),
\begin{equation}
A:\Omega\to\mathbb R_{\ge0},\qquad A(\mathbf r)=|\psi(\mathbf r)|,
\label{eq:rv_amp}
\end{equation}
\begin{equation}
\Phi:\Omega\to[-\pi,\pi),\qquad \Phi(\mathbf r)=\arg\psi(\mathbf r),
\label{eq:rv_phase}
\end{equation}
where $\mathbb R_{\ge0}$ and $[-\pi,\pi)$ define the value spaces over which the amplitude and phase distributions are
supported. The joint and marginal distributions of $(A,\Phi)$ are then the push-forward of $P$ under the map
$\mathbf r\mapsto(A(\mathbf r),\Phi(\mathbf r))$, i.e., the distribution of the values $(A(\mathbf r),\Phi(\mathbf r))$
induced by sampling $\mathbf r\sim P$.

\subsection{Discretized joint distribution and Shannon-type measures (bits)}
For numerical evaluation, we discretize $A\in[0,A_{\max}]$ into $N_A$ bins $\{B_i\}$ and $\Phi$ into $N_\Phi$ circular bins
$\{C_j\}$ on $[-\pi,\pi)$ (identifying $-\pi$ and $\pi$ to avoid edge artifacts). The binned joint probability mass is
computed by a Born-weighted 2D histogram,
\begin{widetext}
\begin{equation}
p_{A,\Phi}(i,j)
=\mathbb{P}(A\in B_i,\ \Phi\in C_j)
=\frac{\sum\nolimits_{r\in\Omega:\,A(r)\in B_i,\ \Phi(r)\in C_j} |\psi(r)|^2}
{\sum\nolimits_{r\in\Omega} |\psi(r)|^2},
\label{eq:pAphi}
\end{equation}
\end{widetext}
which satisfies $p_{A,\Phi}(i,j)\ge0$ and $\sum_{i,j}p_{A,\Phi}(i,j)=1$. The amplitude and phase marginals follow by
summation,
\begin{equation}
p_A(i)=\sum_j p_{A,\Phi}(i,j),\qquad
p_\Phi(j)=\sum_i p_{A,\Phi}(i,j).
\label{eq:marginals_A_Phi}
\end{equation}

In our implementation we used $N_A=300$ and $N_\Phi=256$ bins. The upper amplitude cutoff $A_{\max}$ was chosen globally
for each dataset as
\begin{equation}
A_{\max}=\max_{\text{files}}\ \max_{r\in\Omega}\ |\psi(r)|,
\label{eq:Amax_global}
\end{equation}
so that all parameter-dependent curves are compared on the same amplitude discretization. We verified that the
qualitative trends reported below are stable under reasonable variations of $(N_A,N_\Phi)$.

From these discretized probabilities we compute Shannon-type information measures in bits ($\log_2$). The marginal
entropies are
\begin{equation}
H(A)=-\sum_i p_A(i)\log_2 p_A(i),
\label{eq:HA}
\end{equation}
\begin{equation}
H(\Phi)=-\sum_j p_\Phi(j)\log_2 p_\Phi(j),
\label{eq:HPhi}
\end{equation}
and the joint entropy is
\begin{equation}
H(A,\Phi)=-\sum_{i,j} p_{A,\Phi}(i,j)\log_2 p_{A,\Phi}(i,j).
\label{eq:HAphi}
\end{equation}
We emphasize the conditional entropies,
\begin{equation}
H(A|\Phi)=H(A,\Phi)-H(\Phi),
\label{eq:HA_given_Phi}
\end{equation}
\begin{equation}
H(\Phi|A)=H(A,\Phi)-H(A),
\label{eq:HPhi_given_A}
\end{equation}
and introduce the mutual information,
\begin{equation}
I(A;\Phi)=H(A)-H(A|\Phi)
         =H(\Phi)-H(\Phi|A),
\label{eq:MI_APhi}
\end{equation}
which quantifies the statistical dependence between local amplitude and phase. With these definitions, we next track
$H(A)$, $H(\Phi)$, $H(A,\Phi)$ and $I(A;\Phi)$ across the avoided crossing and relate their behavior to non-Hermitian mode
mixing.

\section{Amplitude--phase signatures of the avoided crossing}
\subsection{Distribution-level reshaping of $p_{A,\Phi}$ across the avoided crossing}
Figure~\ref{Figure-2} provides a distribution-level view of the avoided-crossing (A.C.) response by visualizing the
Born-weighted joint distribution $p_{A,\Phi}(A,\Phi)$ together with the marginals $p_A(A)$ and $p_\Phi(\Phi)$ at
representative parameter points. Panels (a--c) correspond to the marked points A--C on the upper branch (mode~1) in
Fig.~\ref{Figure-1}(a), while panels (d--f) correspond to the marked points D--F on the lower branch (mode~2); in
particular, B and E represent the A.C.\ points for the two branches.

A systematic reshaping occurs at the A.C.\ panels (b,e): the amplitude marginal $p_A(A)$ sharpens while the phase
marginal $p_\Phi(\Phi)$ broadens. This redistribution at the level of the full joint distribution highlights that the
strong-interaction window reorganizes the complex eigenfield in a way that cannot be inferred from intensity-only
diagnostics. The entropy-based analysis below quantifies these distribution-level changes and separates their impact
into the amplitude and phase sectors.

\subsection{Entropic and correlational response: amplitude localization versus phase delocalization}
\begin{figure}
\centering
\includegraphics[width=8.5cm]{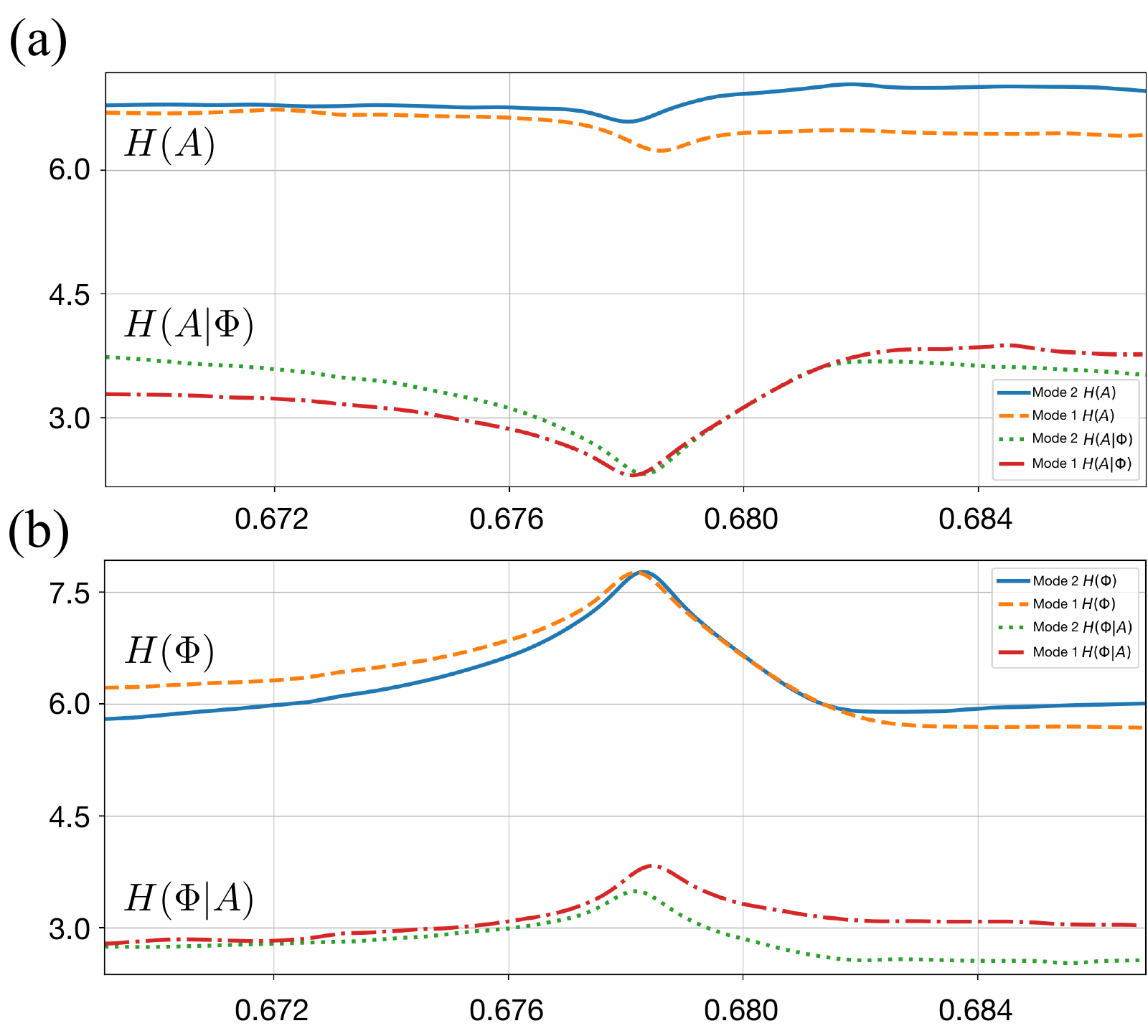}
\caption{\textbf{Amplitude and phase entropies across the avoided crossing.}
(a) Amplitude entropy $H(A)$ and conditional amplitude entropy $H(A|\Phi)$ versus eccentricity $e$ for the two
interacting modes. Both curves exhibit a clear dip in the avoided-crossing (A.C.) window.
(b) Phase entropy $H(\Phi)$ and conditional phase entropy $H(\Phi|A)$ versus $e$. Both curves form a pronounced peak in
the same A.C.\ window. In both sectors, conditioning substantially reduces the entropy, demonstrating strong
statistical dependence between local amplitude and phase under Born-weighted sampling.}
\label{Figure-3}
\end{figure}

The distribution-level changes in Fig.~\ref{Figure-2} translate into sharp information-theoretic signatures when
quantified by the marginal and conditional entropies defined in
Eqs.~\eqref{eq:HA}--\eqref{eq:HPhi_given_A}. In Fig.~\ref{Figure-3} we summarize the A.C.\ response for the two
interacting modes, focusing here on the marginal and conditional measures of the amplitude and phase sectors.

Figure~\ref{Figure-3} quantifies the A.C.\ response using the measures introduced in
Eqs.~\eqref{eq:HA}--\eqref{eq:HPhi_given_A}. The amplitude sector [Fig.~\ref{Figure-3}(a)] exhibits a distinct dip in
both $H(A)$ and $H(A|\Phi)$ in the A.C.\ window, indicating that the amplitude variable $A(r)=|\psi(r)|$ becomes more
localized in its \emph{value space} at the mixing point. This contrasts with the behavior of the intensity-based spatial
entropy $H_P$ (Fig.~\ref{Figure-1}), which increases near the A.C.\ due to real-space hybridization and redistribution
of $|\psi(\mathbf r)|^2$. The two observations are logically consistent because real-space redistribution and value-space
spreading are distinct notions: the A.C.\ can enhance spatial delocalization of the intensity pattern while
simultaneously tightening the distribution of local amplitude values.

The most nontrivial signature appears in the phase sector. In non-Hermitian resonators, $|\psi(r)|^2$ discards the phase
information entirely, so intensity-only observables cannot diagnose how the complex eigenfield reorganizes its phase.
Figure~\ref{Figure-3}(b) shows that the phase entropy $H(\Phi)$ reaches a pronounced maximum in the same A.C.\ window,
demonstrating that $\Phi(r)=\arg\psi(r)$ becomes maximally delocalized in value space at the mixing point. Importantly,
this phase broadening persists under conditioning: $H(\Phi|A)$ also peaks in the A.C.\ window, indicating that the phase
delocalization is not a trivial byproduct of amplitude variations but reflects a genuine restructuring of the complex
field in the strong-interaction regime.

A robust quantitative feature is that conditioning reduces the uncertainty by about a factor of two in both sectors:
$H(A|\Phi)\approx \tfrac{1}{2}H(A)$ and $H(\Phi|A)\approx \tfrac{1}{2}H(\Phi)$, with the strongest contrast near the
A.C.\ window. This shows that amplitude and phase are strongly dependent random variables under the Born-weighted
sampling on $\Omega$, so that specifying one variable significantly constrains the other. We will return to the
two-variable characterization in the next section, where joint measures are discussed together with the integrated
interpretation of the A.C.\ response.

\section{Position-conditioned amplitude--phase dependence and co-information}
\label{sec:Pi_coinformation}

\begin{figure}
\centering
\includegraphics[width=8.5cm]{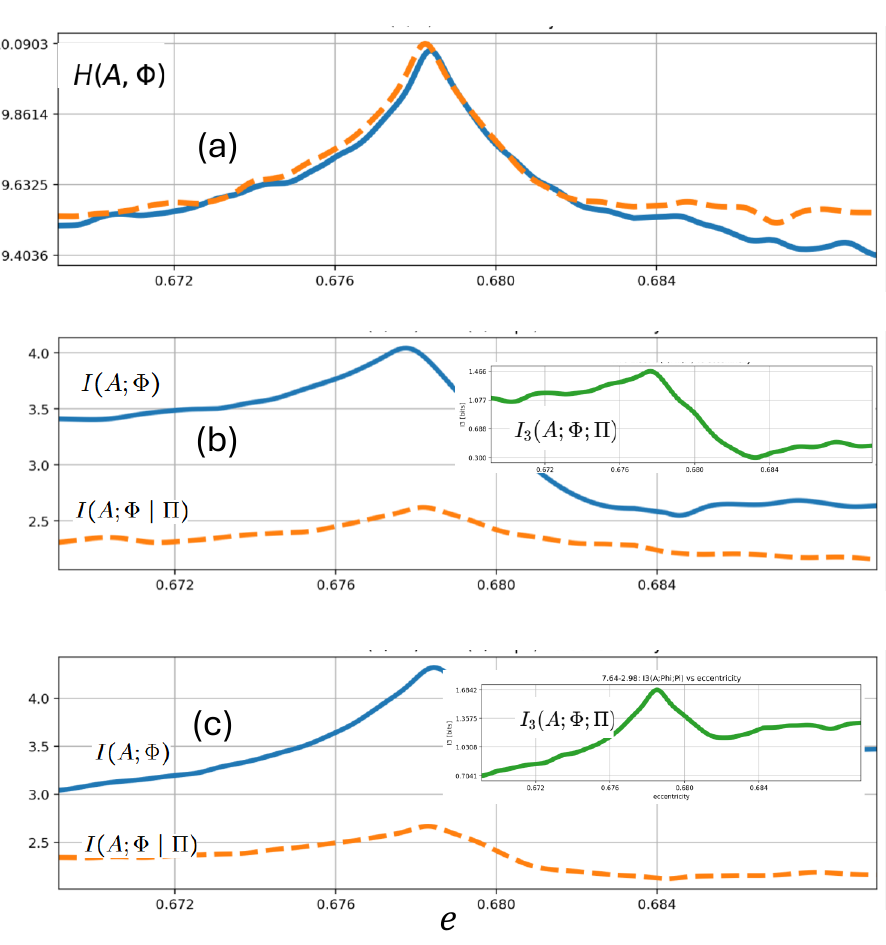}
\caption{\textbf{Joint entropy and position-conditioned amplitude--phase dependence across the avoided crossing.}
(a) Joint entropy $H(A,\Phi)$ versus eccentricity $e$.
(b,c) Global mutual information $I(A;\Phi)$ (solid) and position-conditioned mutual information $I(A;\Phi\mid \Pi)$
(dashed) versus $e$ for the two datasets. Insets plot the co-information
$I_3(A;\Phi;\Pi)=I(A;\Phi)-I(A;\Phi\mid\Pi)$. All information measures use base-2 logarithms and are reported in bits.}
\label{Figure-4}
\end{figure}

\subsection{A coarse position label as a new random variable}
\label{subsec:Pi_rv}
Building on the Born-weighted sampling model on the cavity grid defined in the previous section, we introduce an
additional random variable that encodes \emph{where} in the cavity the field is sampled. We partition the $(x,y)$ domain
into $40\times 40$ spatial bins $\{B_m\}_{m=1}^{M}$ with $M=40^2$, and define the coarse position label
\begin{equation}
\Pi:\Omega\to\{1,\dots,M\},\qquad
\Pi(\mathbf r)=
\begin{cases}
m, & \mathbf r\in B_m .
\end{cases}
\label{eq:Pi_def}
\end{equation}
Here $\mathbf r$ is sampled from the same Born-induced probability mass function $P(\mathbf r)$ as before, so the induced
distribution of $\Pi$ is~\cite{CoverThomas2006Book}
\begin{equation}
p_\Pi(m)=\mathbb{P}(\Pi=m)=\sum_{\mathbf r\in\Omega:\,\Pi(\mathbf r)=m} P(\mathbf r).
\label{eq:pPi}
\end{equation}
Thus, $p_\Pi(m)$ equals the Born-weighted probability mass carried by bin $m$, i.e., the fraction of the mode intensity
contained in that coarse region. Bins with $p_\Pi(m)=0$ carry no Born weight and do not contribute to any
$\Pi$-conditioned quantities; equivalently, conditioning is understood on the support $\{m:\,p_\Pi(m)>0\}$. In this sense,
$\Pi$ provides a physically transparent control variable that separates \emph{within-region} amplitude--phase structure
from \emph{across-region} spatial heterogeneity.

\subsection{Position-conditioned mutual information and co-information}
\label{subsec:CMI_coinf}
We next ask how much of the global amplitude--phase dependence is retained when the coarse spatial label is specified.
The position-conditioned mutual information is defined by~\cite{CoverThomas2006Book}
\begin{equation}
I(A;\Phi\mid\Pi)=H(A\mid\Pi)+H(\Phi\mid\Pi)-H(A,\Phi\mid\Pi).
\label{eq:CMI_APi}
\end{equation}
Equivalently, in terms of the $\Pi$-resolved distributions, it can be written as
\begin{widetext}
\begin{equation}
I(A;\Phi\mid\Pi)
=\sum_{m:\,p_\Pi(m)>0} p_\Pi(m)\sum_{i,j} p_{A,\Phi\mid\Pi}(i,j\mid m)\,
\log_2\!\frac{p_{A,\Phi\mid\Pi}(i,j\mid m)}{p_{A\mid\Pi}(i\mid m)\,p_{\Phi\mid\Pi}(j\mid m)}.
\label{eq:CMI_hist}
\end{equation}
\end{widetext}

To compare the global dependence to the within-bin dependence, we introduce the co-information (interaction information)
\begin{equation}
I_3(A;\Phi;\Pi)=I(A;\Phi)-I(A;\Phi\mid\Pi),
\label{eq:I3_def}
\end{equation}
which is symmetric under permutation of the three variables~\cite{McGill1954Psychometrika,Watanabe1960IBMJRD,Bell2003CoInfoLattice}.
If $I_3(A;\Phi;\Pi)>0$, then conditioning on $\Pi$ \emph{reduces} the amplitude--phase dependence, indicating that a
non-negligible portion of the global $I(A;\Phi)$ is attributable to mixing across spatial bins (spatial heterogeneity at
the coarse scale). If $I_3(A;\Phi;\Pi)<0$, the within-bin dependence is, on average, \emph{stronger} than the globally
mixed dependence, meaning that coarse spatial mixing partially masks locally strong amplitude--phase correlations.


\section{Conclusion}
We introduced a field-level, information-theoretic diagnostic of non-Hermitian mode interaction in an open elliptical
microcavity across an avoided crossing. While the intensity-based spatial entropy peaks in the A.C.\ window, the complex
field shows a complementary split: the amplitude entropy dips and the phase entropy peaks, with conditioning indicating
strong amplitude–phase dependence. A coarse position label further shows that the enhanced global coupling near the
A.C.\ is strongly shaped by spatial heterogeneity across the cavity. These diagnostics provide a compact route to
identifying and comparing strong-interaction regimes in open resonators and can aid mode engineering and control in
wave-based platforms where complex fields are computed or reconstructed (e.g., interferometric phase retrieval and
Wigner/tomographic measurements).

\section{acknowledgement}
This work was supported by the National Research Foundation of Korea (NRF) through a grant funded by the Ministry of Science and ICT (Grants Nos. RS-2023-00211817 and RS-2025-00515537), the Institute for Information \& Communications Technology Promotion (IITP) grant funded by the Korean government (MSIP) (Grants No. RS-2025-02304540), and the National Research Council of Science \& Technology (NST) (Grant No. GTL25011-401). S.L. acknowledges support from the National Research Foundation of Korea (NRF) grants funded by the MSIT (Grant No. RS-2022-NR068791).


\end{document}